\documentclass{webofc}
\usepackage[varg]{txfonts}   
\usepackage{bm}
\usepackage{graphicx}
\usepackage{ascmac}
\usepackage{amsmath,amssymb}
\usepackage{cancel}
\usepackage{braket}
\usepackage{url}
\usepackage[setpagesize=false,hidelinks,pdfusetitle]{hyperref}
\usepackage{mathtools}
\usepackage[normalem]{ulem}  
\usepackage[]{color} 

\renewcommand\sout{\bgroup \color{red} \ULdepth=-.5ex \ULset}

\begin{document}
\title{\texorpdfstring{Hadron-hadron potentials coupled to quark degrees of freedom for $X(3872)$}{Hadron-hadron potentials coupled to quark degrees of freedom for X(3872)}}
%
%

\author{\firstname{Ibuki} \lastname{Terashima}\inst{1}\fnsep\thanks{\email{}terashima-ibuki@ed.tmu.ac.jp} \and
        \firstname{Tetsuo} \lastname{Hyodo}\inst{1}\fnsep\thanks{\email{hyodo@tmu.ac.jp}}
}

\institute{Department of Physics, Tokyo Metropolitan University, Hachioji 192-0397, Japan}

\abstract{%
To study the internal structure of the exotic hadron $X(3872)$, considering the coupled-channel potential between quarks and hadrons is necessary because $X(3872)$ is regarded as a mixture state of $c\bar{c}$ and $D^0\bar{D}^{*0}$. In this work, we construct the hadron-hadron potentials coupled to the quark channel and study the properties of the $D^0\bar{D}^{*0}$ potentials for $X(3872)$. In particular, we compare various local approximations by focusing on the scattering phase shifts.
}
\maketitle
\section{Introduction}
\label{sc:intro}
The study of potentials of hadron interactions reveals their essential mechanism. For example, the origin of the attraction in the nuclear force was explained by Yukawa using $\pi$ exchange potential~\cite{Yukawa:1935xg}. Later, realistic nuclear forces and chiral effective  field theory were developed~\cite{Epelbaum:2008ga,Machleidt:2011zz,Hammer:2019poc}. Now, the nuclear force can also be studied by lattice QCD~\cite{Ishii:2006ec,Aoki:2020bew}.
Meanwhile, the static $Q\bar{Q}$ potentials was found to be useful to study the color confinement in QCD. In fact, the linear quark potential obtained by the strong coupling expansion~\cite{Wilson:1974sk} indicates the confinement. From the viewpoint of phenomenology, the Cornell potential (Coulombic plus linear) is known to reproduce the charmonium spectrum~\cite{Eichten:1974af,Godfrey:1985xj}, which is also confirmed by lattice QCD~\cite{Bali:2000gf}.

Quark potentials and hadron potentials have been studied independently with different degrees of freedom. However, QCD allows the mixing of states with the same quantum numbers. For example, $X(3872)$ is considered to be a mixture of the $c\bar{c}$ and $D\bar{D}^*$ states~\cite{Belle:2003nnu,Takizawa:2012hy}.
As an example of the study of the channel coupling, the string breaking of the static potential has been described by the lattice QCD calculation~\cite{Bali:2005fu}. However, the effect of channel coupling in the hadron potential is not well explored.

In this work, we derive the effective hadron potentials from the coupled-channel problem between quarks and hadrons. We utilize the framework of the hadron-hadron potential with the quark contribution to construct the $D\bar{D}^*$ interaction for $X(3872)$ by numerical calculations. A detailed discussion can be found in Ref.~\cite{Terashima:2023tun}.

\section{Formulation}\label{sc_formu}
We introduce $\ket{c\bar{c}}$ as a quark channel wavefunction and $\ket{D^0\bar{D}^{*0}}$ as a hadron channel wavefunction. To achieve $J^{PC}=1^{++}$ of $X(3872)$, the $c\bar{c}$ channel ($D^0\bar{D}^{*0}$ channel) is combined in the $^3{\rm P}_1$ state (s wave)\footnote{Hereafter, ``$D^0\bar{D}^{*0}$'' stands for the abbreviation of the linear combination $(\ket{D^0\bar{D}^{*0}}+\ket{D^{*0}\bar{D}^{0}})/\sqrt{2}$.} 
In this study, we follow the Feshbach's method~\cite{Feshbach:1958nx,Feshbach:1962ut} to describe the effective interaction in the coupled-channel system for $X(3872)$. The Hamiltonian $H$ that couples the $c\bar{c}$ channel with the $D^0\bar{D}^{*0}$ channel by the transition potential $V^t$ and the coupled-channel Schr\"odinger equation are given by
\begin{align}\label{eq_2ch_hamiltonian_X}
    H=
    \begin{pmatrix}
        T^{c\bar{c}} & 0                           \\
        0            & T^{D^0\bar{D}^{*0}}+ \Delta
    \end{pmatrix}
    +
    \begin{pmatrix}
        V^{c\bar{c}} & V^t \\
        V^t          & 0
    \end{pmatrix}, \quad
        H     \begin{pmatrix}
        \ket{c\bar{c}} \\
        \ket{D^0\bar{D}^{*0}}
    \end{pmatrix} =  E  \begin{pmatrix}
        \ket{c\bar{c}} \\
        \ket{D^0\bar{D}^{*0}}
    \end{pmatrix},
\end{align}
where $T^{c\bar{c}}$ and $T^{D^0\bar{D}^{*0}}$ are the kinetic energies of the quark and hadron channels, respectively, $V^{c\bar{c}}$ is the confinement potential for quarks, and $\Delta$ is the threshold energy of the hadron channel. 

By eliminating the quark channel, the effective potential between hadrons are obtained as
\begin{math}\label{eq_Veff_h}
    V^{D^0\bar{D}^{*0}}_\mathrm{{eff}}(E)=V^{D^0\bar{D}^{*0}}+V^tG^{c\bar{c}}(E)V^t 
\end{math}
with the Green's function for the quark channel 
\begin{math}
    G^{c\bar{c}}(E) = (E-(T^{c\bar{c}} + V^{c\bar{c}}))^{-1}
\end{math}
, which satisfies the single-channel Schr\"odinger equation
\begin{math}\label{eq_sch_h}
     \left[T^{D^0\bar{D}^{*0}}+ \Delta+ V^{D^0\bar{D}^{*0}}_\mathrm{{eff}}(E)\right] \ket{D^0\bar{D}^{*0}}= E \ket{D^0\bar{D}^{*0}}.
\end{math}

To obtain the coordinate space representation $\bra{\bm{r}}V^{D^0\bar{D}^{*0}}_\mathrm{{eff}}(E)\ket{\bm{r}^{\prime}}=V_\mathrm{ eff}^{D^0\bar{D}^{*0}}(\boldsymbol{r},\boldsymbol{r'},E) $, we solve the
Schr\"{o}dinger equation for the quark channel in the absence of the channel coupling,
\begin{math}
    (T^{c\bar{c}} + V^{c\bar{c}})\ket{\phi_0} =E\ket{\phi_0}.
\end{math}
Since  $V^{c\bar{c}}$ is the confinement potential, there are no scattering solutions, and only the discrete eigenstates exist.
Among the complete set of the discrete eigenstates of the $c\bar{c}$ potential in the $^3{\rm P}_1$ channel, we pick up the contribution from the $\chi_{c1}(2P)$ state which locates slightly above the $D^0\bar{D}^{*0}$ threshold. Here we adopt the Yukawa-type transition form factor $\bra{\phi_{0}}V^{t}\ket{\bm{r}}=g_{0}e^{-\mu r}/r$ with $\mu$ being the cutoff.
We then obtain the nonlocal effective $D^0\bar{D}^{*0}$ potential coupled with the $\phi_{0}$ state as
\begin{align}\label{eq_VDDeff}
    V_\mathrm{ eff}^{D^0\bar{D}^{*0}}(\boldsymbol{r},\boldsymbol{r'},E)   = \frac{g_0^2}{E-E_0}\frac{e^{-\mu r}}{r}\frac{e^{-\mu r'}}{r'},
\end{align}
where $E_{0}$ is the energy of $\phi_{0}$ relative to the threshold of the $D^0\bar{D}^{*0}$ channel and $g_{0}$ is the coupling constant which is determined to reproduce the mass of $X(3872)$ as in Ref.~\cite{Terashima:2023tun}.
Due to the channel coupling, we see that the effective potential $ V^{D^0\bar{D}^{*0}}_\mathrm{ eff}(\bm{r}, \bm{r'}, E)$ is nonlocal and energy dependent, and also it diverges at the quark channel eigenenergy $E=E_0
$.

Because the physical properties of the nonlocal potentials are not clear, 
we introduce the formal derivative expansion and the derivative expansion by the HAL QCD method based on Ref.~\cite{Aoki:2021ahj}, which are the approximation methods of converting the nonlocal potential to a local one. 
In this work, we generalize the methods in Ref.~\cite{Aoki:2021ahj} for the energy dependent nonlocal potential as in Eq.~\eqref{eq_VDDeff}.
The formal derivative expansion is the method of expanding nonlocal potentials in powers of the derivatives by the Taylor expansion. We obtain the lowest order local potential from Eq.~\eqref{eq_VDDeff} by the formal derivative expansion as
\begin{align}
    V^\mathrm{{formal}}(r,E)                & =\frac{g_0^2}{E-E_0}\frac{4\pi}{\mu^2}\frac{e^{-\mu r}}{r}.\label{eq_Vformalyukawa}
\end{align}
In the HAL QCD method, we assume that the wavefunction from the nonlocal potential satisfies the Schr\"odinger equation with a local potential, and then invert the equation to get the local potential. To apply this method to the energy dependent nonlocal potential \eqref{eq_VDDeff}, we calculate the scattering wavefunction $\psi_k(r)$ and the phase shift $\delta$ at a fixed momentum $k$. Thanks to the separable form of Eq.~\eqref{eq_VDDeff}, the wavefunction and the phase shift are obtained analytically. For instance, the explicit expression of the phase shift $\delta(k)$ is given as 
\begin{align}\label{eq_exact_delta}
     & k \cot\delta(k) 
     = \frac{4\pi m \mu g_0^2-\mu^4 E_0}{8\pi m g_0^2}   +\frac{8\pi m^2 g_0^2-\mu^5+4m\mu^3E_0}{16\pi m^2\mu g_0^2}k^2   +\frac{-\mu^2+m E_0}{8\pi m g_0^2}k^4-\frac{1}{16\pi m^2 g_0^2}k^6.
\end{align}
Using the wavefunction and the phase shift, we obtain the local approximated potential by the HAL QCD method at the lowest order as
\begin{align}\label{eq_VHAL_kaiseki}
          & V^\mathrm{ HAL}(r;k_0) = \frac{k_0^2}{2m}  +  \frac{-k_0^2\sin \left[k_0 r+\delta(k_0)\right]-\mu^2\sin\delta(k_0) e^{-\mu r}}{2m\{\sin\left[k_0 r+\delta(k_0)\right]-\sin\delta(k_0) e^{-\mu r}\}},
\end{align}
where $k_{0}$ is the momentum at which the wavefunction $\psi_{k_0}$ is calculated.
We note that the local potential by the HAL QCD method~\eqref{eq_VHAL_kaiseki} is energy independent, but depends on this momentum $k_0$. The Schr\"{o}dinger equation with this potential is
$\left[-\nabla^2/(2m) +V^\mathrm{ HAL}(r;k_0) \right]\psi(\bm{r}) =E\psi(\bm{r})$.
At $E = (k_0)^2/(2m)$, we obtain $\psi=\psi_{k_0}$ and the exact phase shift of the nonlocal potential is reproduced. However, at $E\neq ({k_0})^2/2m $, the exact wavefunctions are in general not obtained.

\section{Numerical results}

We numerically study the local $D^{0}\bar{D}^{*0}$ potentials, which have a shallow bound state $X(3872)$ with the binding energy $B=40$ keV.
We set $E_0 \simeq0.078$ GeV from the quark model~\cite{Godfrey:1985xj} and the cutoff $\mu=0.14$ GeV from the $\pi$ exchange. 

\begin{figure}[tb]
\centering
\includegraphics[width=6.5cm,clip]{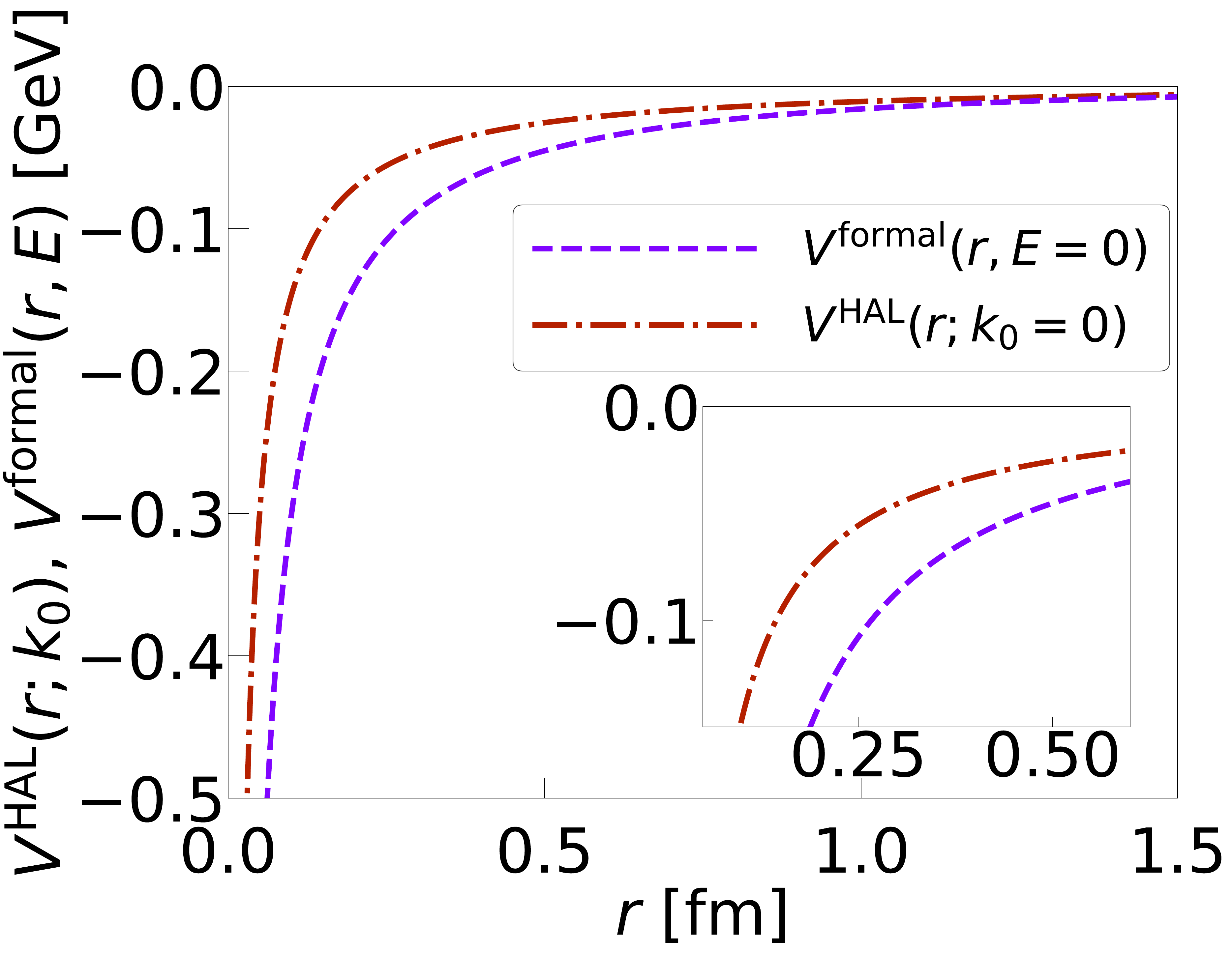}
\includegraphics[width=6.2cm,clip]{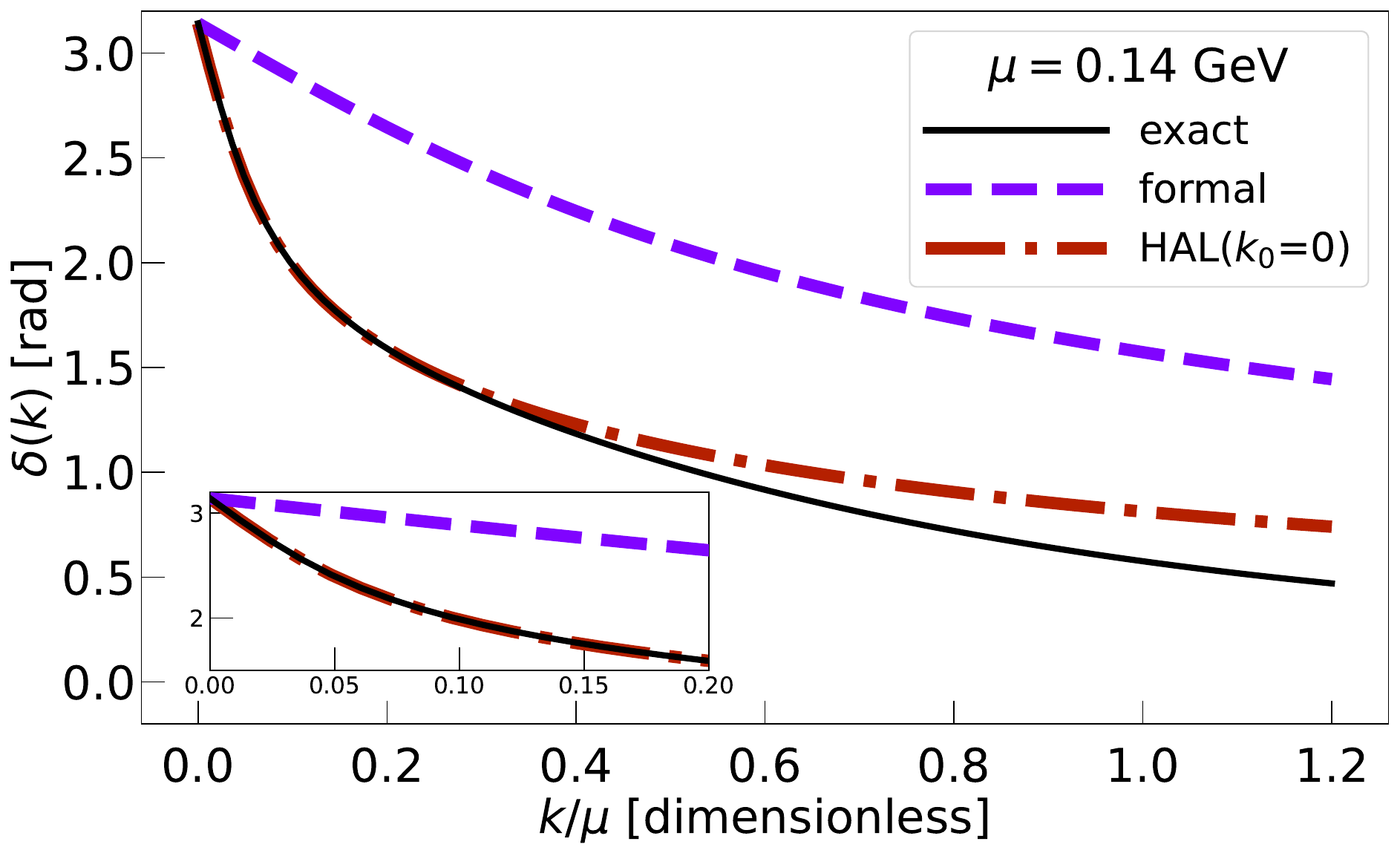}
\caption{[left] The comparison of the local potentials by the formal derivative expansion $V^\mathrm{{formal}}(r,E=0)$ (dashed line) and by the HAL QCD method $V^{\mathrm{ HAL}}(r;k_0=0)$ (dashed-dotted line) as functions of the relative distance $r$. [right] The phase shifts from the local potentials by the formal derivative expansion (dashed-dashed line) and by the HAL QCD method (dotted line) as functions of the dimensionless momentum $k/\mu$ with the cutoff $\mu=0.14$ GeV in comparison with the exact phase shift (solid line).}
\label{fig}
\end{figure}

We compare the local potential by the formal derivative expansion with $E=0$ (dashed line) and the one by the HAL QCD method with $k_0=0$ (dashed-dotted line) in the left panel of Fig.~\ref{fig}.
We find the quantitative difference of the potentials, even though both are constructed from the same nonlocal potential~\eqref{eq_VDDeff}. In particular, the difference is pronounced in the short distance region of $r\lesssim 0.4 ~\mathrm{ fm}$. 

We next investigate how the difference of the local potentials affects on the phase shift $\delta(k)$.
We show the phase shifts $\delta$ by the formal derivative expansion (dashed line) and by the HAL QCD method (dashed-dotted line) as functions of the dimensionless momentum $k/\mu$ in the right panel of Fig.~\ref{fig}.
By comparing the dashed line with the dashed-dotted line, we find that the difference in potentials seen in the left panel affects the phase shifts quantitatively.
We also show the exact phase shift $\delta$ (solid line) by the original nonlocal potential~\eqref{eq_exact_delta} in Fig.~\ref{fig}.
We find that the exact phase shift is better reproduced by the HAL QCD method than the formal derivative expansion.
In particular, the potential by the HAL QCD method works well in the small $k$ region, indicating that the scattering length defined by the slope of the phase shift at $k=0$ is also reproduced.

\section{Summary}\label{sc_summary}
In this paper, we have discussed the properties of the effective potentials with the channel coupling of the quark and hadron degrees of freedom. We show that the effective single-channel potential is nonlocal and energy dependent, reflecting the channel coupling effect. We examine the local approximation methods of the obtained nonlocal effective potential using the model of $X(3872)$ to extract the physical mechanism of the interaction. In this study, we introduce two approximation methods: the formal derivative expansion and the HAL QCD method. 

The approximated local potentials and the scattering phase shifts from the potentials are computed numerically. We show the quantitative deviation of the obtained local potentials and phase shifts for different approximation methods. 
It is found that the HAL QCD method works better than the formal derivative expansion, even for the energy-dependent nonlocal potential.

This work has been supported in part by the Grants-in-Aid for Scientific Research by JSPS (Grant numbers
JP22K03637, 
JP19H05150, 
JP18H05402), 
and by JST, the establishment of university fellowships towards the creation of science technology innovation, Grant Number JPMJFS2139.

\end{document}